\documentclass[10pt,preprint,a4paper]{aastex}

\usepackage{graphics,epsf}
\usepackage{amsmath}                
\usepackage{amsfonts}               
\usepackage{amssymb}                
\usepackage{epsfig}                 

\def \cm{~\rm{cm}}

\def \K{~\rm{K}}
\def \g{~\rm{g}}

\def \erg{~\rm{erg}}

\def \yr{~\rm{yr}}

\def \days{~\rm{day}}

\title{THE ENERGY SOURCE OF INTERMEDIATE LUMINOSITY OPTICAL TRANSIENTS\footnote{Presented at the
Intermediate-Luminosity Red Transients meeting, STScI, Baltimore, USA, June 28-30, 2011.} }

\author{Noam Soker\altaffilmark{2} and Amit Kashi\altaffilmark{2}}

\altaffiltext{2}{Department of Physics, Technion -- Israel Institute of
Technology, Haifa 32000 Israel;  soker@physics.technion.ac.il; kashia@physics.technion.ac.il.}

\begin{document}

\begin{abstract}
We argue that transient systems with luminosity between novae and supernovae (SNe) are powered by gravitational energy
of mass accreted onto, or a companion merges with, a main-sequence star.
These transient events are termed Intermediate-Luminosity Optical Transients (ILOTs; other terms in use are
Intermediate-Luminosity Red Transients and Red Novae).
We show that despite the wide range of $10^{45}$--$10^{50} \erg$, the typical energy released by ILOTs can be
expressed as a function of fundamental variables $F_I(\hbar, c, G, m_e, m_N, kT_i)$,
the planck constant, speed of light, gravitational constant, electron mass, neutron mass, and
ignition temperature of hydrogen. This expression explains why ILOTs are located between SNe and novae
with respect to their total energy. We also put an upper limit on the power (luminosity) of ILOTs, which
explains their lower luminosity than SNe.
\end{abstract}

\section{INTRODUCTION}
\label{sec:intro}

The total energy released in supernova (SN) explosions are about six orders of magnitude higher than the total energy
released in novae explosions. By \emph{total energy} we refer here to the sum of kinetic energy of the
ejected material and the total radiated energy. By \emph{available energy} we refer to the total energy
that is released in the powering process.
Observations slowly filling the gap between novae and SNe
(e.g., Barbary et al. 2009; Berger et al. 2009, 2011; Bond et al. 2009; Kulkarni et al. 2007a,b;
Kulkarni \& Kasliwal 2009; Ofek et. al. 2008; Rau et al. 2007; Kasliwal et al. 2011; Monard 2008; Prieto et al. 2008;
Nakano 2008; Mason et al. 2010 ; Mould et al. 1990; Pastorello et al. 2010; Smith et al. 2011; Wesson et al. 2009).
In our presentations at the \emph{Intermediate-Luminosity Red Transients} meeting
we argued that the total energy of these transients hints on the basic energy source (`engine') powering their eruptions.
We term these outbursts Intermediate-Luminosity Optical Transients (ILOTs; other terms in use are
Intermediate-Luminosity Red Transients and Red Novae).

In this summary of our presentations we show that
in the same way as the total energy of SNe and novae can be explained from fundamental quantities,
so is the case for ILOTs. For that we summarize in short the expressions for SNe and novae energies,
and then turn to discuss ILOTs.
We propose that ILOTs are powered by accreting mass from a companion onto a main-sequence (MS) star.
The companion can either survive the mass transfer process, or merge with the primary in a process termed mergerburst.

\section{THE ENERGY-TIME DIAGRAM (ETD)}
\label{sec:ETD}

\begin{figure*}
\resizebox{1.0\textwidth}{!}{\includegraphics{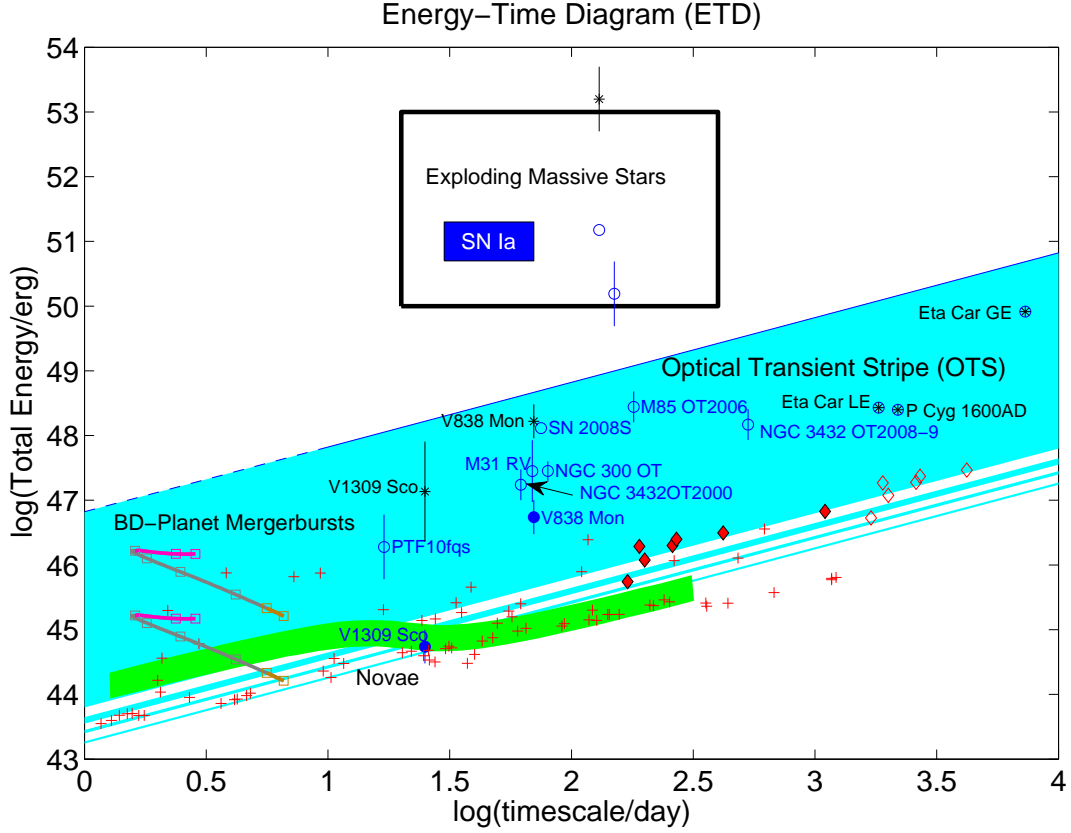}}
\caption{\footnotesize{Observed transient events on the Energy-Time Diagram (ETD; taken from Kashi \& Soker 2011).
Also plotted are the calculated location of the predicted Brown dwarf (BD)-planet mergerbursts.
Blue empty circles represent the total (radiated plus kinetic) energy
of the observed transients $E_{\rm{tot}}$ as a function of the
duration $t$ of their eruptions.
The Optical Transient Stripe (OTS), is a more or less constant luminosity region in the ETD.
It is populated by accretion powered events such as ILOTs (including mergerbursts),
major LBV eruptions, and the predicted BD-planets mergerbursts.
Blue filled circles represent observed transients that are
mergerbursts. The green line represents nova models
computed using luminosity and duration from della Valle \& Livio (1995).
Nova models from Yaron et al. (2005) are marked with red
crosses, and models from Shara et al. (2011) are represented with diamonds.
The total energy does not include the energy which is
deposited in lifting the envelope that does not escape from the
star. Where a model exists to calculate the gravitational
energy released by the accreted mass (the available energy), it is
marked by a black asterisk above or overlapping the blue circle.
For detail on the BD-Planet mergerbursts see Bear et al. (2011).}}
\label{fig:totEvst}
\end{figure*}

One way to present the different objects is the energy-time diagram (ETD; at the workshop,
M. Kasliwal and S. Kulkarni presented the classification of ILOTs on the luminosity-time diagram; see Kulkarni et al. 2007a).
The ETD (Figure \ref{fig:totEvst}) presents the \emph{total energy} of the transients, radiated plus kinetic,
as a function of the duration of their eruption, defined as a drop of $3$ magnitudes in the V-band.
However, as the goal is eventually to understand the physical processes of the eruptions, when a model for
an event exists, or when there is more information from observations, we calculate the
\emph{available energy}, i.e., total gravitational energy available for the event. Namely, the gravitational energy that could
have been released if all the mass is accreted by the accreting star.

The upper-right region of the OTS is occupied by major LBV eruptions,
while the lower-left region is occupied by ILOTs.
The lower most part of the OTS is predicted to be occupied by mergerbursts between a planet and a brown dwarf (BD)
(Bear et al. 2011).
In this process the planet is shredded into a disk, and the accretion lead to an outburst. The destruction of
a component in a binary system and transforming it to an accretion disk is an extreme
case of mass transfer processes in binary systems.

For major LBV eruptions the available energy is equal to the radiated plus kinetic energy, as there is no
inflated envelope.
We estimate the available energy for some ILOTs.
However, as stated, for most ILOTs the observations and models are not yet detailed enough to perform this estimate,
and we can only present the estimated radiated plus kinetic energy.

\section{SUPERNOVAE}
\label{sec:SNe}

The available energy of $\sim 10^{53} \erg$ (including neutrinos) in the explosion of core-collapse (CC)
SNe can be understood from fundamental quantities.
The energy is the binding energy of a neutron star (NS). The mass of the neutron star is
about the Chandrasekhar mass that is given by
$M_{\rm NS} \simeq M_{\rm Ch} \simeq (\hbar c / G)^{3/2} / (m_n)^2$, where $m_n \simeq m_H$ is the neutron mass.
The NS radius is given from basic relation of the pressure of degenerate neutron gas
$R_{\rm NS} \simeq \hbar^2 /(G m_n^{8/3} M_{\rm NS}^{1/3})$.
Therefore, the total available energy is
\begin{equation}
E_{\rm CC-avaible} \simeq \frac{G M_{\rm NS}^2}{R_{\rm NS}} =
F_{\nu}(\hbar, c, G, m_N) \simeq 10^{53} \erg,
\label{eq-ccavailable}
\end{equation}
where $F_{\nu}$ is a function of the fundamental parameters.
(The numerical values are given up to a factor of $\sim 10^{1/2}$, as we are not interested in the exact values,
but only in orders of magnitude.)

The total explosion energy (kinetic + radiation) of CC SNe is about the binding energy of the star, which is
basically the binding energy of the core. This is about equal to the binding energy of a white dwarf (WD), which
is also the total explosion energy of SNe Ia. The relevant mass is the Chandrasekhar mass.
The expression for the typical radius of a WD, or of the core of a CC SN, includes the mass of the electron $m_e$ as well
as the other fundamental quantities
$R_{\rm WD} \simeq \hbar^2 /(G m_e m_n^{5/3} M_{\rm WD}^{1/3})$
The total energy of CC and type Ia SNe is therefore
\begin{equation}
E_{\rm SNe} \simeq \frac{G M_{\rm WD}^2}{R_{\rm WD}} =
F_{\rm SN}(\hbar, c, G, m_e, m_N) \simeq 10^{51} \erg.
\label{eq-cctotal}
\end{equation}

\section{NOVAE}
\label{sec:novae}

The energy released by novae can also be understood from fundamental quantities, although with a larger scatter than
the expressions relevant to SNe.
A hydrogen rich gas is accumulated onto a WD, and is ignited when the temperature at the base of
the accreted layer reaches the hydrogen nuclear ignition temperature of $T_i \simeq 10^7 \K$.
We demand that the gas at the base of the accreted layer is degenerate at this temperature.
The constraint on the density of the degenerate electron gas comes from quantum mechanics and
reads $(\rho/m_H)^{2/3} \sim kT_im_e/\hbar^2$
and gives $\rho_i \sim 10^3 \g \cm^{-3}$. At this density the thermal pressure is about equal to the
pressure due to degenerate electrons gas.
Equating the hydrostatic pressure of the accreted layer to the gas pressure at temperature $T_i$ and density
$\rho_i$ gives the mass of the accreted layer. This can be cast into the form
\begin{equation}
\left( \frac{M_{\rm acc}} {M_{\rm WD}} \right)_{\rm ignition} \sim
\frac {\rho_i}{\rho_{\rm WD}} \frac{T_i}{T_{\rm virial-WD}} \equiv F_N(\hbar, c, G, m_e, m_N, kT_i),
\label{eq-maccwd}
\end{equation}
where $\rho_{\rm WD}$ is the average density of the WD, and $T_{\rm virial-WD}$ is the virial temperature of the WD.
We have ${\rho_i}/\rho_{\rm WD} \sim 10^{-3}$ and ${T_i}/T_{\rm virial-WD} \sim 10^{-2}$,
from which we derive $M_{\rm acc} \sim 10^{-5} M_{\rm WD} \simeq 10^{-5} \rm{M_\odot}$.
The efficiency of hydrogen burning is several percents, accounting for novae typical total energy of
$\sim 10^{44}$--$10^{46} \erg$.

\section{ILOTs}
\label{sec:ILOTs}

We here argue that the typical available energy of ILOTs can also be accounted for by fundamental quantities.
The available energy in these events (see Fig. 1) is $E_{\rm ILOT} \sim 10^{45}$--$10^{50} \erg$.
We show that this energy, despite the wide range, is a typical binding energy of main-sequence (MS)
stars and envelopes of Asymptotic Giant Branch (AGB) and Red Giant Branch (RGB) stars.

The requirement that hydrogen burns in the core of MS stars set their virial temperature
at $T_{\rm virial-MS} \simeq {T_i}$. In turn, this determines the gravitational potential on the stellar surface
\begin{equation}
\frac{G M_{\rm MS}} {R_{MS}} \simeq  \frac{k T_i}{\mu m_H} \simeq 3 \times 10^{48} \erg ~ \rm{M_\odot}^{-1},
\label{eq-MS1}
\end{equation}
where $\mu m_H$ is the mean mass per particle. The potential can be $\sim 3$ times as high for massive stars,
as the secondary star in $\eta$ Carinae, and 3 times lower for brown dwarfs.

The amount of mass accreted in our model (Kashi et al. 2010; Kashi \& Soker 2011; Bear et al. 2011) might vary a lot:
It can be $\sim$~several$\times \rm{M_\odot}$ in the case of the Great Eruption of $\eta$ Carinae
(Gomez et al. 2006, 2009; Smith et al. 2003; Smith \& Ferland 2007; Smith \& Owocki 2006),
and down to $\sim 10^{-3} \rm{M_\odot}$ in a case where a planet is destructed on the surface of a brown dwarf
(Bear et al. 2011). This gives the large variation in the energy.
However, the gross typical mass is that of a MS star, which is about the solar mass.
This mass is also determined from fundamental parameters: The upper limit is set by demanding that
radiation pressure does not destroy the star, while the lower mass is determined by the brown dwarf limit
where degenerate material prevents the contraction and heating-up the core.
Therefore, the typical mass that a companion can transfer to a MS companion, either through merger or
mass transfer from a red giant is $M_{\rm acc} \sim 1 \rm{M_\odot}$.
The available energy is
\begin{equation}
E_{\rm available-ILOT} \simeq \frac{G M_{\rm MS}} {R_{MS}} M_{\rm acc} \simeq  \frac{k T_i}{\mu m_H} \rm{M_\odot}
\equiv F_I(\hbar, c, G, m_e, m_N, kT_i) \simeq 3 \times 10^{48} \erg .
\label{eq-MS2}
\end{equation}
This, we suggest, explains most of the events populating the
the optical transient stripe (OTS) between novae and SNe.
The total energy might be less than the available energy, as some of the energy goes to uplift envelope mass,
rather than to the kinetic energy of the ejected mass and to radiation.

Interestingly, The binding energy of the envelopes of massive AGB stars (extreme AGB), those with masses of $\sim 5$--$10 \rm{M_\odot}$,
is $\sim 10^{48} \erg$. This might raise the possibility that events related to the envelopes,
such as major instability of the envelope or ejection of the envelope, might be also located in that region.
For example, the ejection of the envelope of an AGB star of $\sim 5 \rm{M_\odot}$ requires energy of $\sim 10^{48} \erg$.
If this is done via a common envelope phase, it can be an explosive event with the break out of a shock wave
(Kashi \& Soker 2011), and the release of this typical amount of anergy. The `engine' that supplies the energy
is the gravitational energy of the core and a companion that end at an orbital separation of $\sim 1 R_\odot$,
bringing us back to an equation similar to equation (\ref{eq-MS2}).
The shock break-out in this case might not have the power that is required in the scenario of Kochanek (2011)
for dust destruction.
However, an accreting companion might supply the required ionizing radiation that can destroy the dust
in systems like NGC~300OT (Bond et al. 2009).

\section{TIME SCALE}
\label{sec:time}

The time scales of SNe, novae, and ILOTs are determines by different processes. For ILOTs, it is basically
the mass transfer time scale. Here we set a lower limit on this time scale.

The average total gravitational power is the average accretion rate times the potential well of the accreting star
\begin{equation}
L_G=\frac{G M_a \dot{M_a}}{R_a},
\label{eq:L}
\end{equation}
where $M_a$ and $R_a$ are the mass and radius of the accretor, and $\dot{M_a}$ is the mass accretion rate.
In the binary model discussed here, accreted mass is likely to form an accretion disk or an accretion belt.
The accretion time must be longer than the viscosity time scale for the accreted mass to lose its angular momentum.
According to Dubus et al. (2001) the viscous timescale is
\begin{equation}
t_{\rm{visc}}
\simeq \frac{R_a^2}{\nu}
\simeq 73
\left(\frac{\alpha}{0.1}\right)^{-1}
\left(\frac{H/R_a}{0.1}\right)^{-1}
\left(\frac{C_s/v_\phi}{0.1}\right)^{-1}
\left(\frac{R_a}{5 \rm{R_{\odot}}}\right)^{3/2}
\left(\frac{M_a}{8 \rm{M_{\odot}}}\right)^{-1/2} \days, 
\label{eq:tvisc1}
\end{equation}
where $\nu$ is the viscosity of the disk, $H$ is the thickness of the disk,
$C_s$ is the sound speed and $v_\phi$ is the Keplerian velocity.
We scale $M_a$ and $R_a$ in equation (\ref{eq:tvisc1}) according to the parameters of V838~Mon (Tylenda 2005).
For these parameters the viscous to Keplerian times ratio is $\chi \equiv t_{\rm{visc}}/t_K \simeq 160$.

The accreted mass is determined by the details of the binary interaction process, and differs from object to object.
We scale it by $M_{\rm{acc}} = \eta_a M_a$.
Based on the modelled systems (V~1309~Sco, Tylenda et al. 2011; V838~Mon; $\eta$ Car) this mass fraction is
$\eta_a \lesssim 0.1$ with a large variation.
The value of $\eta_a \lesssim 0.1$ can be understood as follows.
If the MS (or slightly off-MS) star collides with a star and tidally disrupts it, as in the model for V838~Mon (Soker \& Tylenda 2003; Tylenda \& Soker 2006),
the destructed star is likely to be of much lower mass than the accretor $M_{\rm{acc}} \lesssim M_b \lesssim 0.3 M_a$.
Another possibility is that an evolved star loses a huge amount of mass.
In that case it is possible that the accretor will gain only a small fraction of the ejected mass,
as in the scenario for the Great Eruption of $\eta$ Carinae (Kashi \& Soker 2010).
Here again we expect $M_{\rm{acc}} \lesssim 0.1 M_a$.

The viscous time scale gives an upper limit on the accretion rate
\begin{equation}
\dot{M} < \frac{\eta_a M_a}{t_{\rm{visc}}}
\simeq 4
\left(\frac{\eta_a}{0.1}\right)
\left(\frac{\alpha}{0.1}\right)
\left(\frac{H/R_a}{0.1}\right)
\left(\frac{C_s/v_\phi}{0.1}\right)
\left(\frac{R_a}{5 \rm{R_{\odot}}}\right)^{-3/2}
\left(\frac{M_a}{8 \rm{M_{\odot}}}\right)^{3/2} ~\rm{M_{\odot} \yr^{-1}}. 
\label{eq:dotM}
\end{equation}
The maximum gravitational power is therefore
\begin{equation}
L_G < L_{\rm{max}} = \frac{GM_a\dot{M_a}}{R_a}
\simeq 7.7 \times 10^{41}
\left(\frac{\eta_a}{0.1}\right)
\left(\frac{\chi}{160}\right)^{-1}
\left(\frac{R_a}{5 \rm{R_{\odot}}}\right)^{5/2}
\left(\frac{M_a}{8 \rm{M_{\odot}}}\right)^{-5/2}
\erg ~\rm{s^{-1}},
\label{eq:Lmax}
\end{equation}
where we replaced the parameters of the viscous time scale with the ratio of viscous to Keplerian time $\chi$.
Equation (\ref{eq:Lmax}) sets the upper bound on the OTS in the ETD, plotted as blue line at the upper edge of the OTS.
We note that the location of this line may change if the accretion efficiency $\eta$ is different
and/or the stellar parameters of the accreting star are different.
For most of the ILOTs the accretion efficiency is lower, hence they are located below this line,
giving rise to the relatively large width of the OTS.
The uncertainty in $\eta_a$ is large and in extreme cases may be even $>1$.
Therefore, on rare occasions we expect to find objects slightly above the upper line drawn in the figure.

\section{SUMMARY}
\label{sec:summary}

We here take the view that the systems populating the peak-luminosity region, or total energy region,
between novae and SNe are not a collection of different kinds of objects. Rather, we think that the
typical total energy (kinetic + radiated) of these events (termed ILOTs) points to one basic energy source:
The gravitational energy of mass accreted onto, or merged with, a main-sequence star.
The energy scale of ILOTs comes from the basic properties of
main-sequence stars, namely, that they burn hydrogen.
First, the potential on their surface is related to the hydrogen ignition temperature by
equation (\ref{eq-MS1}). Second, the available energy for a major event is the mass a companion
can supply times the potential. The typical mass in a merger process is that of another main-sequence star.
In a case of a mass transfer from the envelope of an evolved star a similar mass is available.
In many cases less mass is transferred, and the total energy is lower even (section \ref{sec:time}).

The power of the basic engine that leads to ILOTs is limited by viscosity that limits the duration of the process
from below. This in turn constrains the power of the event from above.
This time is derived in section \ref{sec:time} and drawn by the thick blue line on Figure \ref{fig:totEvst}.

\end{document}